# Single-Molecule Circuits with Well-Defined Molecular Conductance

Latha Venkataraman[1,4], Jennifer E. Klare[2,4], Iris W. Tam[2,4], Colin Nuckolls[2,4],

Mark S. Hybertsen[3,4], Michael L. Steigerwald[2]

[1]Department of Physics, [2]Department of Chemistry,

[3]Department of Applied Physics and Applied Mathematics, and

[4]Center for Electron Transport in Molecular Nanostructures

Columbia University, New York, New York

AUTHOR EMAIL ADDRESS: latha@phys.columbia.edu

**Abstract**: We measure the conductance of amine terminated molecules by breaking Au point-contacts in a molecular solution at room temperature.  We find that the variability of the observed conductance for the diamine molecule-Au junctions is much less than the variability for diisonitrile and dithiol–Au junctions. This narrow distribution enables unambiguous conductance measurements of single molecules. For an alkane diamine series with 2-8 carbon atoms in the hydrocarbon chain our results show a systematic trend in the conductance from which we extract a tunneling decay constant of $0.91 \pm 0.03$ per methylene group. We hypothesize that the diamine link binds preferentially to under-coordinated Au atoms in the junction. This is supported by density functional theory based calculations that show the amine binding to a gold adatom with sufficient angular flexibility for easy junction formation but well-defined electronic coupling of the N lone pair to the Au. Therefore, the amine linkage leads to well defined conductance measurements of a single molecule junction in a statistical study.

---

A statistical comparison of single molecule devices is a prerequisite for understanding the nanoscopic details of circuitry built from molecules. Here we detail a new and versatile end-group/metal pairing between gold and amines forming molecular junctions with a narrow conductance distribution that allows for a meaningful comparison of numerous measurements of single molecules in electrical circuits. The majority of candidates for end-group/metal pairings for molecular electronics come from studies of self-assembled monolayers such as thiolated molecules on gold surfaces. Self-Assembled Monolayers (SAMs) on metals.[1] For example, the thiols and gold are the most prevalent metal-molecule binding motif. However, many of the driving forces for thiols to form stable SAMs, such as the variability in the bonding between an end-group and a metal or favorable van der Waals interaction between the molecules, are orthogonal to the properties needed for reliable electrical measurements on single molecules. For thiols on gold, the force required to break the Au-S bond is larger than that for the Au-Au bond,[2-4] leading to grain migration, pitting of electrodes,[5,6] and metal island formation that complicate interpretation of single-molecule measurements.[7] In this study we show that molecular junctions, formed using amine endgroups with gold point contacts,[2,8,9] allow for the conductance to be measured of a diverse set of molecules. The tunneling decay constant of a series of alkane diamines is $0.91 \pm 0.03$ per methylene group. Density functional theory (DFT) calculations indicate that the lone pair of the amines bind preferentially to under-coordinated Au structures such as an ad-atom and provide a well defined electronic coupling between the N lone pair and the Au.

We measure the molecular conductance by repeatedly forming and breaking Au point contacts in the presence of molecules with a modified scanning tunneling microscope (STM). The STM operates in ambient conditions at room temperature and the junctions were broken in a dilute 1mM solution of the molecules in 1,2,4-trichlorobenzene.[10] This system relies on the trapping of one or more molecules between the ends of an Au point contact that is being broken by pulling the electrodes apart. The current was recorded at a fixed bias voltage while the electrodes were pulled apart, to generate conductance traces. Thousands of curves were collected to allow for detailed statistical analysis. In addition, to ensure that each measurement started from a different initial atomic configuration, the electrodes were pulled apart only after being brought into contact with the Au surface, indicated by a conductance greater than a few $G_0$, where $G_0 = 2e^2/h$, the quantum of conductance.

When a clean Au contact is pulled and broken, its conductance decreases in a stepwise fashion (inset, Figure 1). The steps are at integral multiples of $G_0$, indicating that the cross-section of the contact is reduced down to that of a few and eventually a single atomic chain.[11] A conductance histogram constructed from 1000 consecutive traces shows clear peaks around 1, 2, and $3 \times G_0$ (Figure 1). When the single atom chain is broken in the absence of molecules, the conductance either decreases exponentially with the electrode displacement due to tunneling across the gap between the two Au electrodes or it drops from $G_0$ to below our experimental detection limit possibly due to the broken ends snapping back as the contacts relax (Figure 2a, yellow traces).

The conductance histograms of three differently substituted aromatics, 1,4-benzenedithiol, 1,4-benzenediisonitrile, and 1,4-benzenediamine, are compared in Figure 2b. For each of the molecules, a large fraction of the individual traces show additional steps at conductance levels below $G_0$, as shown in Figure 2a. Conductance histograms are constructed from more than 3000 consecutively measured conductance traces without any data selection or processing so as to include all possible molecule/electrode conformations in the statistical analysis.[12] In comparison to the data for the dithiol or the diisonitrile, the conductance histogram for 1,4-benzenediamine is strikingly well-defined. Indeed, within our experimental accuracy, the histogram differs from that of Au only for conductance between 0.003 $G_0$ and 0.015 $G_0$. This implies that the additional steps occur only within this conductance range. From a Gaussian fit to this peak (inset Figure 2b), we deduce the most prevalent conductance value for the molecule to be 0.0064 ± 0.0004 $G_0$. A Gaussian cannot be fit to the diisonitrile data although a broad peak is visible in Figure 2b between 0.001 $G_0$ and 0.1 $G_0$. The conductance for the dithiol occurs over a wide range. The resulting histogram is diffused showing no well-defined trend in the conductance.[13] (See supporting information for other control experiments.)

The broadening of the conductance histograms of the diisonitriles and dithiols is not unexpected. Thiols are known to bind to gold in a variety of motifs, e.g. atop and three-fold hollow sites. The binding of sulfur over atop and three-fold hollow sites has been calculated to yield a difference of up to a factor of 3 in the conductance.[14] Moreover, the dithiols may contain oligomeric molecules formed through oxidative disulfide formation. Isonitriles[15] suffer from similar complications. They are known to oligomerize easily[16] and also have a variation in their binding to the surface gold atoms.[17]

We have measured the conductance of a series of alkanes with amine end-groups that progresses from 1,2-ethylenediamine to 1,8-octanediamine. The conductance histogram for each alkane (Figure 3a) displays a prominent primary peak.[18] We determined the conductance of these molecules from the center position of a Gaussian fit to the primary peaks.[19] Figure 3b shows these conductance values plotted against the number of methylene groups on a semi-log scale along with an exponential fit to the data. The clear exponential dependence provides additional evidence that the peak in the histograms corresponds to the conductance of a single molecule and not an ensemble. The tunneling decay constant, β, determined from the slope is 0.91 ±0.03 per methylene group or equivalently, 0.77 ± 0.03 per Å. This β value, though slightly smaller, is comparable to that of alkanedithiols determined by different methods (0.8±0.08 per Å,[2] 0.83 ± 0.04 per Å,[20] or 0.94± 0.06 per Å).[2, 20, 21] This small difference could be due to a difference in the alignment of the molecule HOMO and LUMO relative to the Au Fermi level as compared with thiolated molecules.[22] Extending this fit to the origin, we can infer a contact resistance of ~ 430 kΩ for two amine-Au bonds. This is a factor of 10 larger than the value obtained for alkanethiols from conducting atomic force microscopy (AFM) measurements,[23] although this difference could be due to the fact that our experiments are not done under large compressive forces unlike the conducting AFM experiments.

What is different about the amine-Au link chemistry in this experiment? First, the Au surfaces are not smooth. It is well known from coordination chemistry that simple ligands such as amines bind more strongly to coordinatively-unsaturated metal centers;[24] amines are not known to form stable SAMs on Au but are used to passivate Au nanoparticles.[25] Second, to a first approximation, Au may be considered a locally isotropic, one-electron metal with an intact d-shell. The amines are strong bases and very weak acids; hence they bond to metals as two-electron donors and rarely as covalently-bonded amides (i.e., M-NR$_2$). We suggest, therefore, that the bonding between Au and the amines is a simple delocalization of the lone-pair of electrons from the amine-nitrogen to coordinatively-unsaturated surface Au atoms. There is just this one mode of Au-amine bonding. As implied by the locally isotropic electronic structure around the Au binding site, the Au-amine bond will not be strongly directional so the molecular junction formation is relatively unconstrained by the link structure and can easily form. Finally, the frontier electronic orbital responsible for electronic transport has the local character of an antibonding hybrid of the Au s-orbital and the N lone pair. This will follow the Au-N bond and be relatively insensitive to the local structure. Together, these attributes explain the observation of sharp molecular conductance signatures for amine-Au linkages.

To investigate the amine-gold interaction further, we have performed DFT based calculations within the generalized gradient approximation (GGA)[26] both for 1,4-butanediamine bonded to Au clusters[27, 28] as well as to periodic Au slabs.[29,30] (See supporting information for details.) We find that on a flat Au(111) surface, the 1,4-butanediamine does not bind near the a-top site in an upright configuration, but does binds to an Au adatom with an estimated binding energy of 0.5 eV. This suggests a scenario for the junction structure illustrated in Figure 4a. To explore the flexibility of this binding, the energy surface has been probed as a function of the Au-N-C bond angle, the Au-Au-N bond angle and rigid rotation around the axis going through the adatom using an $Au_5$ cluster to model each electrode and keeping all other bond lengths and angles fixed. As shown in Figure 4b, the energy cost to flex the Au-N-C angle by ±15 degrees is about 0.2 eV (counting contributions from both contacts). Interestingly, the Au-Au-N angle is a much softer degree of freedom, especially to the wide angle side. The energy cost for axial rotation is quite small (<0.1 eV).

In Figure 4c, we show, the frontier electronic state of the 1,4-butanediamine bonded to an adatom responsible for the tunneling conductance. The tunneling matrix element can be estimated from the splitting of the appropriate frontier orbital energies in cluster analogues of the junction. The conductance is proportional to the square of the energy splitting, in the weak tunneling limit. Analysis of this model for the alkane series with 2,4,6 and 8 methylene units with the diamines bonded to Au adatoms gives a tunneling decay constant β of 0.94/methylene, in good agreement with our experimental results and expectations from analysis of evanescent states for the alkane polymer.[22] An important factor for our experiments is the variation in the tunneling coupling with the geometric details of the contact. We find that overall, across the geometry scans studied, the coupling ranges from ½ to 2 times nominal, comparable with the peak widths found in the experimental data.

If indeed, the amine-gold link is through the N lone pair, we can attenuate the binding by adding steric bulk to the nitrogen center. Figure 5 shows conductance histograms of measurements of 1,3-propanediamine, *N,N'*dimethylpropane1,3-diamine and *N,N,N',N'* Tetramethylpropane1,3–diamine (see Figure 5 for structures). Only the 1,3-propanediamine histogram has a clear peak whereas histogram counts occur over a wide range of conductance values for the other molecules. If only one methyl is present on each nitrogen, there is a small but significant increase around 0.003 $G_0$, close to the peak position of 1,3-propanediamine. No clear peak is seen in the tetramethylated propanediamine. This confirms the hypothesis that the N lone pair participates in the binding and conduction through the molecule.

In conclusion, we have measured the statistical distributions of the conductance of amine terminated molecules using the point-contact method and compared them with the distributions of conductance for thiol and isonitrile terminated molecules. The conductance of such an amine-Au bond does not vary significantly from junction to junction due to the isotropy of the N lone pair to Au electronic coupling across the range of metal-molecule configurations. This enables the systematics and statistics of the conductance to be investigated across a range of molecules of interest for molecular electronics. To find similar trends in the histograms of other endgroups, such as thiols on gold, is difficult and can lead to counter intuitive results. For example, histograms of 1,4-benzene dimethanethiol in gold junctions shows a significantly lower conductance than the fully saturated 1,6-hexanedithiol.[2, 31] Here, we find a clear molecular signature with amine-terminated molecules using the point contact technique due to the specificity with which amines bind to under-coordinated gold atoms in the junction.

We thank Horst Stormer, Philip Kim, Ronald Breslow and James Yardley for useful discussions. This work was supported primarily by the Nanoscale Science and Engineering Initiative of the National Science Foundation under NSF Award Number CHE-0117752 and by the New York State Office of Science, Technology, and Academic Research (NYSTAR). CN thanks the Camille Dreyfus Teacher Scholar Program (2004) and the Alfred P. Sloan Fellowship Program (2004). J.E.K. thanks the American Chemical Society Division of Organic Chemistry for the graduate fellowship sponsored by Organic Syntheses. I.W.T. thanks the NSF for the pre-doctoral fellowship. M.L.S. thanks the Material Research Science and Engineering Center program of the NSF under award number DMR-0213574.

**Figures:**

**Figure 1:** Conductance histogram constructed from 1000 traces measured while breaking Au point contacts in the absence of molecules on a linear scale. The bias is 25mV and bin size is $10^{-4}$ $G_0$. The inset shows 5 traces at arbitrary x-axis offsets.

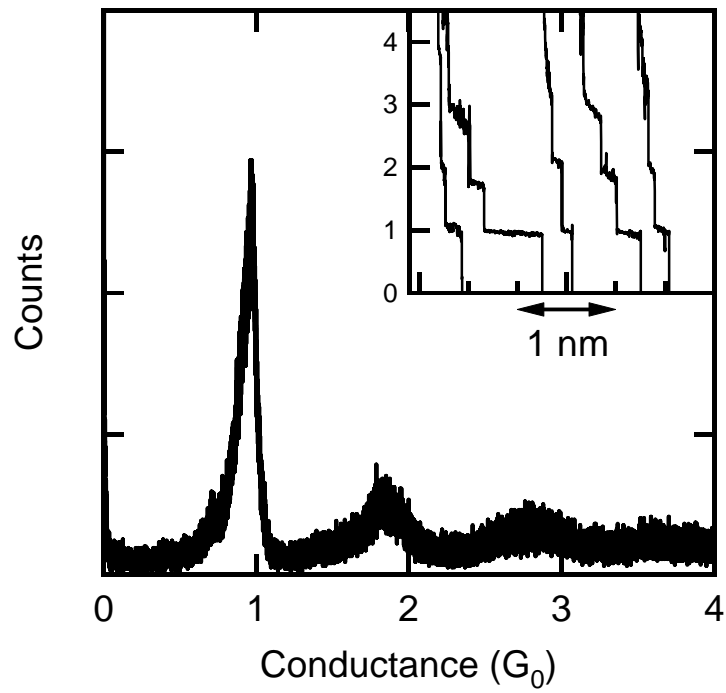

**Figure 2:** (a) Sample conductance traces measured without molecules (yellow) and with 1,4-Benzenediamine (blue), 1,4-benzenedithiol (red), and 1,4-benzenediisonitrile (green) shown on a semi-log plot. All data measured at 25 mV bias although no bias dependence was found up to 250 mV. (b) Conductance histograms constructed from over 3000 traces measured in the presence of 1,4-benzenediamine (blue), 1,4-benzenedithiol (red), and 1,4-benzenediisonitrile (green) shown on a log-log plot. Control histogram of Au without molecules is also shown (yellow). Histograms are normalized by the number of traces used to construct histograms. Inset: Same data on a linear plot showing Gaussian fit to the peak (black curve). Bin size is $10^{-4}$ $G_0$.

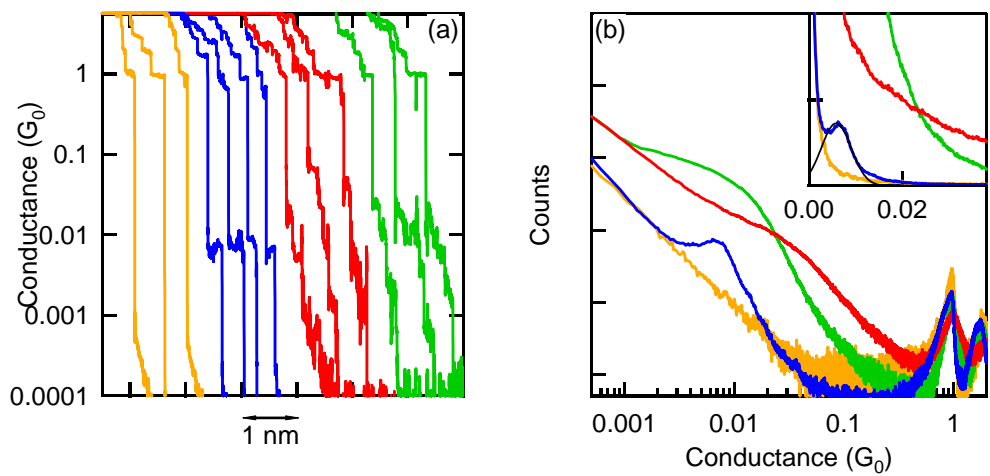

**Figure 3:** (a) Conductance histograms constructed from consecutive traces measured in the presence of alkanediamines with different number of methylene groups in the chain shown on a log-log plot. Histograms are normalized by the number of traces used to construct histograms. Solid arrows point to major peak in the data and dashed arrows point to faint secondary peak possibly due to two molecules bridging the gap. Inset: 1,4-butanediamine histogram showing Gaussian fit to the primary and secondary peak. The bin size varies between $10^{-5}$ and $10^{-6}$ $G_0$ for different traces and all data are measured at 25 mV. Data is offset vertically for clarity. (b) Center position obtained from a Gaussian fit to the peak in the conductance histograms plotted against the number of methylene groups in the alkane chain shown on a semi-log plot.

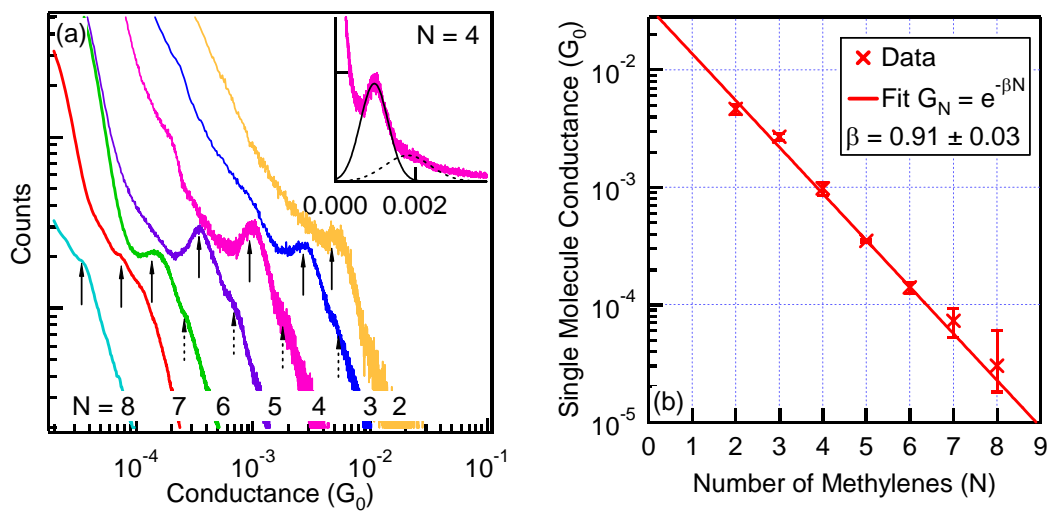

**Figure 4:** (a) Illustration of the hypothesized placement of the 1,4-butanediamine in the junction with the N atoms binding to under-coordinated Au, adatoms here. (b) Calculated change in the total energy versus bonding angle at the adatom, using a cluster analogue representing the adatoms at the contacts with $Au_5$ clusters. (c) Isosurface plot illustrating the frontier orbitals responsible for the tunneling conductance.

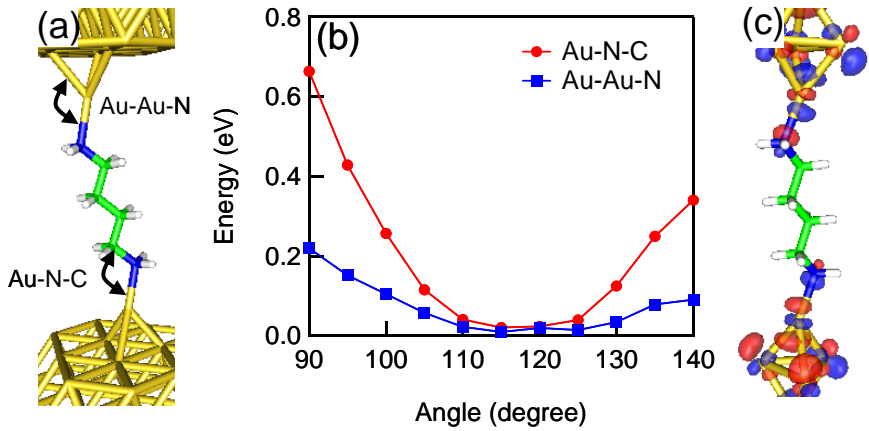

**Figure 5:** (a) Conductance histograms constructed from consecutive traces measured in the presence of 1,3-propanediamine (blue), *N,N'*dimethylpropane1,3diamine (green), and *N,N,N',N'*Tetramethylpropane1,3diamine (red). Control histogram in Au is also shown (yellow). The bin size is $10^{-5}$ $G_0$ and all data are measured at 25 mV. Data is offset vertically as shown by color arrows for clarity.

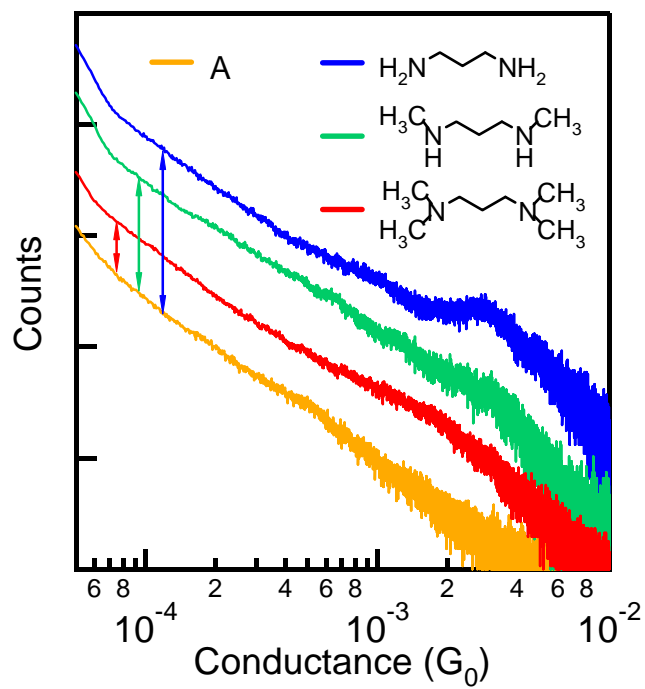

18. For the shorter alkanes, a faint secondary peak is also visible at twice the conductance value of the primary peak. This could be due to the trapping of two molecules in parallel between the electrodes.

19. For 1,7 Heptanediamine and 1,8 Octanediamine data, we subtract the gold histogram from our data prior to fitting a peak as the conductance of these molecules is close to our detection limit. This adds a large error to the conductance value, as shown in Figure 3b.